\documentclass[twoside]{article}

\input ibvs2.sty
\input epsf.sty

\newcommand{\ltsimeq}{\,\raise 0.3 ex\hbox{$ < $}\kern -0.75 em 
 \lower 0.7 ex\hbox{$\sim$}\,} 

\begin{document}
\IBVShead{6003}{11 November 2011}

\IBVStitletl{Implications of the non-detection of X-ray emission}{from HD\,149427}

\IBVSauth{Stute, M.$^1$; Luna, G.J.M.$^{2,3}$} 

\IBVSinst{Institute for Astronomy and Astrophysics, Section Computational
Physics, Eberhard Karls University T\"ubingen, Auf der Morgenstelle 10, 72076 
T\"ubingen, Germany}
\IBVSinst{Instituto de Ciencias Astronomicas, de la Tierra y del Espacio 
(ICATE),  Av.Espana Sur 1512, J5402DSP, San Juan, Argentina}
\IBVSinst{Harvard-Smithsonian Center for Astrophysics, 60 Garden St.
MS 15, Cambridge, MA, 02138, USA}

\SIMBADobjAlias{HD\,149427}{PC 11}
\IBVStyp{ symbiotic star }
\IBVSkey{binaries: symbiotic, stars: individual (HD\,149427),
stars: white dwarfs, X-rays: stars, ISM: jets and outflows}

\IBVSabs{HD\,149427 is a very enigmatic object. It has been classified either}
\IBVSabs{as a planetary nebula or as a D'-type symbiotic star. Its distance is}
\IBVSabs{also highly uncertain. Furthermore, HD\,149427 is a potential jet }
\IBVSabs{source.}
\IBVSabs{We report the non-detection of X-ray emission from HD\,149427 and}
\IBVSabs{explore the implications to its nature.}
\IBVSabs{We observed the object with {{\em XMM-Newton}} with an effective}
\IBVSabs{exposure time of 33.5 ks.}
\IBVSabs{The upper limit for the flux of the X-ray emission in the soft band}
\IBVSabs{($<2$ keV) is 10$^{-15}$ erg s$^{-1}$ cm$^{-2}$, while in the hard band}
\IBVSabs{($>2$ keV) it is about 10$^{-14}$ erg s$^{-1}$ cm$^{-2}$.}
\IBVSabs{We discuss the implication of our results in light of the possible} 
\IBVSabs{natures of HD\,149427 -- being a planetary nebula or a symbiotic star,}
\IBVSabs{close or very distant. The derived upper limits on the mass accretion}
\IBVSabs{rate of the white dwarf are untypical for symbiotic stars and may} 
\IBVSabs{favor the picture of HD\,149427 being a young PN. HD\,149427 might be}
\IBVSabs{a symbiotic star in hibernation -- if a symbiotic star at all.}
\IBVSabs{We estimate the possible mass-loss rate and kinetic luminosity of the}
\IBVSabs{jet and find no contradiction with our upper limit of soft X-ray}
\IBVSabs{emission. Therefore the jet may be still present but it was too faint}
\IBVSabs{to be detected via soft X-ray emission.}

\begintext

HD\,149427 (PC 11 = IRAS 16336-5536 = PN G 331.1-05.7) was first noticed to be 
a peculiar object by Webster (1966). Its exact nature is not known, several 
scenarios exist in the literature. 

The object is listed in the Planetary Nebula (PN) catalogue of Peimbert \& 
Costero (1961) as PC 11, in the catalogue of Southern PNe of Henize (1967) as 
Hen 2-172 and in the catalogue of Henize (1976) as Hen 3-1223 based on its 
H$\alpha$ emission. Henize (1976) noted that the emission-line spectrum is that 
of a PN but with a continuum of an F star. Gutierrez-Moreno et al. (1987) 
systematically investigated HD\,149427 using optical photographic, spectro-
photometric and spectroscopic observations. They suggested that it is indeed a 
young peculiar PN. Gutierrez-Moreno et al. (1995) confirmed this with a diagram 
using the ratios of [OIII]$\lambda$4363/H$\gamma$ and 
[OIII]$\lambda$5007/H$\beta$. Parthasarathy et al. (2000) suggested that the 
central star is a close binary system with an early-F dwarf companion based on 
UV variations. Webster (1966) classified the object as a Symbiotic Star (SyS) 
based on its spectrum. Using NIR observations, Glass \& Webster (1973) 
classified HD\,149427 as SyS with a yellow supergiant as cool component, 
however, their NIR observations could not reliably identify its nature. 
Allen \& Glass (1974) confirmed this result with new NIR observation, but 
pointed out the high density and high excitation in the nebulosity. Allen 
(1982) listed it as a D'-type SyS. 

The distance is also very uncertain. Milne \& Aller (1975) detected 5 GHz radio 
emission and estimated the distance $>6.7$ kpc based on its radio flux. 
Milne (1979) revised the 5 GHz catalogue of PNe and listed a radio distance of 
$>10.5$ kpc. However, it has been shown that a nearby radio source may be 
confused with HD\,149427 (Wright \& Allen 1978). Maciel (1984) used a 
relationship between nebular ionized mass and radius and estimated a distance 
of $<6.2$ kpc. Gutierrez-Moreno et al. (1987) estimated a distance of about 3 
kpc. Kenny (1995) used radio observations for determining 
the distance of about 5 kpc. Gutierrez-Moreno \& Moreno (1998) measured the 
spectral type of the late component of the central star and, using NIR colors 
of Glass \& Webster (1973), determined a small distance of 420 pc assuming a 
luminosity class V. Assuming that the cool component is, however, a giant of 
luminosity class III, this estimated distance had to be increased to 850 pc 
(Gutierrez-Moreno \& Moreno 1998). Tajitsu \& Tamura (1998) derived a distance 
of 9 kpc using IRAS four-band fluxes. Phillips (2001) used the observed radial 
velocities of PNe and the galactic rotation curve for measuring the distance to 
3.27 kpc. Pereira et al. (2010) analized high resolution spectra of the 
late-type companion and found $\log g$ and $T$ values implying a distance of 
9.94 kpc. However, using the luminosity of the white dwarf in the system, the 
distance of about 10 kpc would result in a extremely large white dwarf radius 
of 0.14 $R_\odot$ (Pereira et al. 2010).

HD\,149427 received our attention since it is a member of the list of symbiotic 
stars with possible jet detections compiled by Brocksopp et al. (2004). 
Brocksopp et al. (2003) found three peaks and extended emission in HST WFPC2 
snapshot images taken in July 1999 with F502N and F656N filters corresponding 
to the central source and knots at distances of about 2'' and 12''. Radio 
observations show extended emission in the same direction as the peaks in the 
HST images (Brocksopp et al. 2003). Gutierrez-Moreno \& Moreno (1998) found 
evidence for jet-like [OIII] emission moving away from the central nebula with 
a velocity of 120 km s$^{-1}$.

X-ray observations provide a direct probe of the two most important components 
of jet-driving systems: the bow and internal shocks of the jet emitting soft 
X-rays and the central parts of the jet engine, where gas is being accreted to 
power the jet, leading to hard X-ray emission. Currently R\,Aqr (Kellogg et al. 
2001, 2007) and CH\,Cyg (Galloway \& Sokoloski 2004; Karovska et al. 2007, 2010)
are the only two jets from symbiotic stars that have been resolved in X-rays. 
All objects with jets, detected in other wavelength, when observed in X-rays, 
show soft components ($<2$ keV; R Aqr, CH Cyg, MWC 560, RS Oph, AG Dra, Z And, 
V1329\,Cyg). The three objects CH Cyg, R Aqr, MWC 560 also emit hard components 
(Mukai et al. 2007; Nichols et al. 2007; Stute \& Sahai 2009). Z And showed hard
emission in one of three observations only (Sokoloski et al. 2006).

The paper is organized as follows: in Section \ref{sec_obs}, we show details of 
the observations and the analysis of the data. After that we describe the 
results in Section \ref{sec_res}. We end with a discussion and conclusions in 
Sections \ref{sec_dis} and \ref{sec_concl}. 

\section{Observation and analysis} \label{sec_obs}

We observed the field of HD\,149427 with {\em XMM-Newton} in 2009 
(Table \ref{Tbl_obs}) using the EPIC instrument operated in full window mode 
and with the medium thickness filter. Simultaneously, we used the Optical 
Monitor OM. All the data reduction was performed using the Science Analysis 
Software (\textsc{SAS}) software package\footnote{http://xmm.vilspa.esa.es/}
version 8.0. We removed events at periods with high background levels from the 
pipeline products selecting events with pattern 0--4 (only single and double 
events) for the pn and pattern 0--12 for the MOS, respectively, and applying 
the filter {FLAG=0}. The resulting exposure time after these steps is 33.5 ks.

\begin{table}[!htb]
\caption{Observations on September 01/02, 2009 (ObsID: 0604920201)} 
\label{Tbl_obs}
\centering
\begin{tabular}{lllll}
\hline\hline
Instrument & Filter & Duration & UT Start & UT Stop \\
\hline
pn   & Medium & 68430                 & 19:05:32 & 14:06:02 \\
MOS1 & Medium & 69504                 & 18:43:12 & 14:01:36 \\
MOS2 & Medium & 69515                 & 18:43:11 & 14:01:46 \\
\hline
OM   & U      & 10$\times$ $\sim$1500 & 18:51:36 & 23:49:51 \\
     & UVW1   & 10$\times$ $\sim$1500 & 00:40:11 & 06:08:27 \\
     & UVM2   & 15$\times$ $\sim$1500 & 06:13:48 & 15:02:12 \\ 
\hline
\end{tabular}
\end{table}

\section{Results} \label{sec_res}

\subsection{Images}

All three X-ray images with the EPIC pn, MOS1 and MOS2 cameras show no detection
of emission above background levels centered on HD\,149427 using the 
coordinates from \textsc{SIMBAD} (Fig. 1, white circles). In all
bands, however, a faint source is visible which is about 26'' away from the 
nominal position of HD\,149427. According to Kerber et al. (2008), the proper 
motion of HD\,149427 is of the order of a few mas in RA and about 20 mas in Dec,
thus too small for explaining this discrepancy. In order to check, whether 
possible pointing errors arose during our XMM observations, we created images in
the soft and hard band for all three cameras, ran a source detection with 
\textsc{edetect} and finally identified the closest 11 objects listed by 
\textsc{SIMBAD} in our mosaic. Fig. 2 shows the mosaic 
including our detected X-ray sources (white circles), the reference objects 
from \textsc{SIMBAD} (green circles) and also the closest source from the ROSAT 
faint source all-sky survey (red circle). While the pointing uncertainties in 
the ROSAT mission are obvious, we can clearly identify three objects, for which 
the \textsc{SIMBAD} position agrees perfectly with the position found in our 
source detection run (white arrows). Thesefore we can exclude any pointing error
and indeed report the non-detection of HD\,149427 in X-rays.

\IBVSfig{20cm}{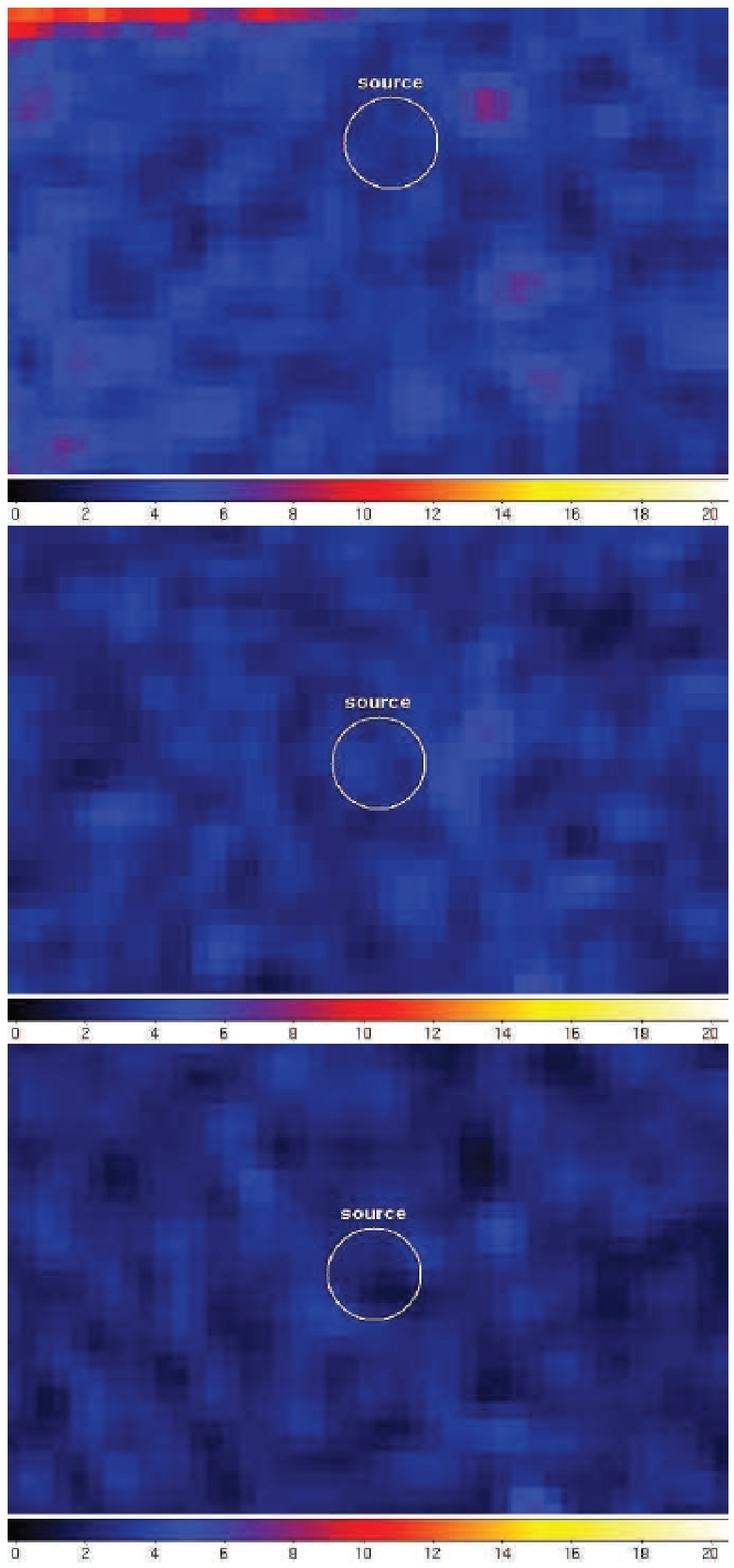}{EPIC images of the region around HD\,149427; top: pn, middle: MOS1, bottom: MOS2; colors show the number of counts.}

\IBVSfig{10cm}{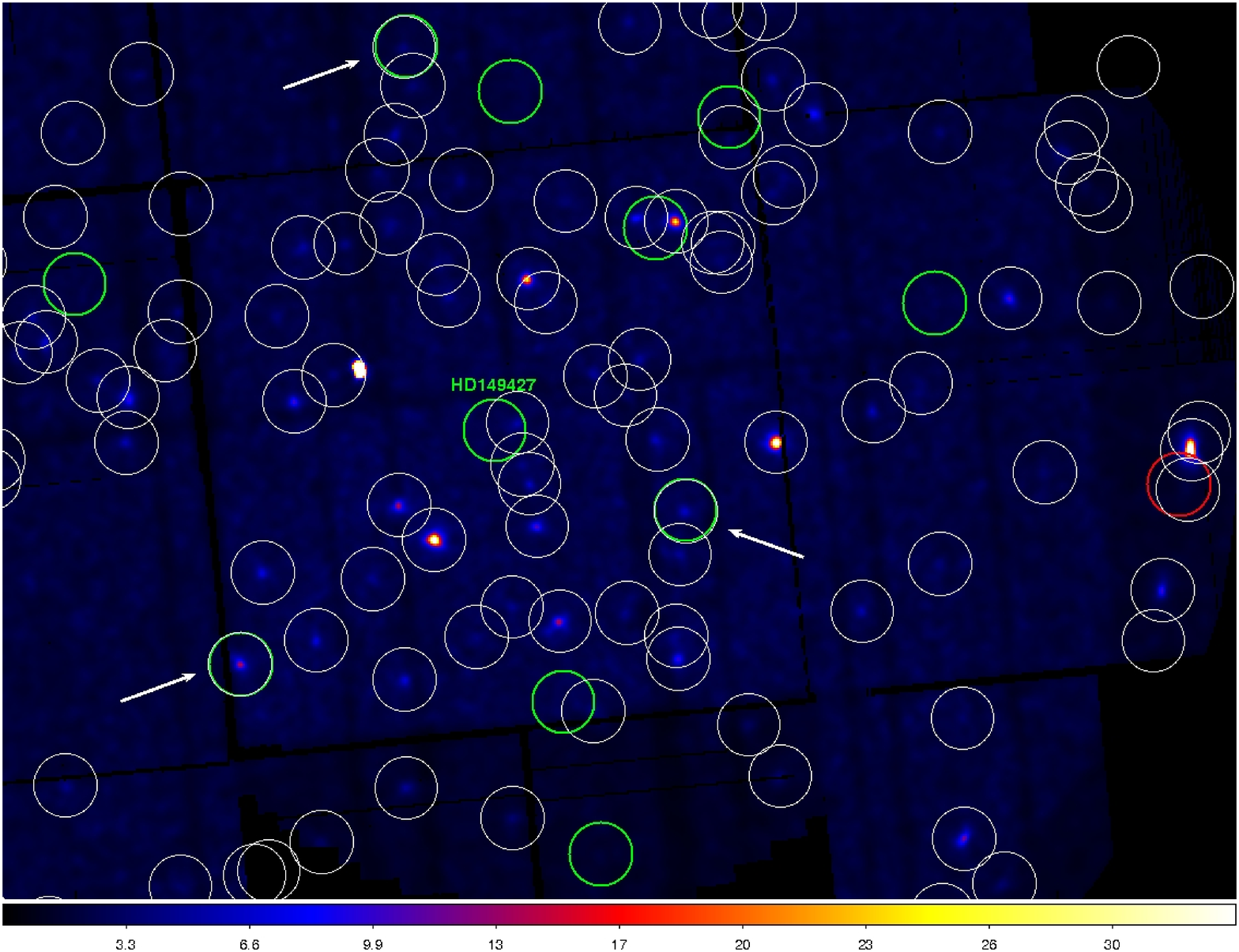}{Mosaic of EPIC images of the region around HD\,149427 including our detected X-ray sources (white circles), the reference objects from \textsc{SIMBAD} (green circles) and also the closest source from the ROSAT faint source all-sky survey (red circle). We can clearly identify three objects, for which the \textsc{SIMBAD} position agrees perfectly with the position found in our source detection run (white arrows) and thus can exclude any pointing error and indeed report the non-detection of HD\,149427 in X-rays.}

Contrary to the X-ray band, HD\,149427 has been detected in the three optical
bands used. The average magnitudes in the optical filters U, UVW1 and UVM2 are 
12.7, 13.5 and 14.8, respectively. The slope of the spectral energy 
distribution indicates that the optical flux is dominated by nebular emission 
(Fig. 3). The flux levels of blackbody emission corresponding to 
a white dwarf with a radius of $8\times10^8$ cm and with an effective 
temperature of 105000 K at a distance of 850 pc (Gutierrez-Moreno \& Moreno 
1998) are lower than the measured fluxes.

\IBVSfig{10cm}{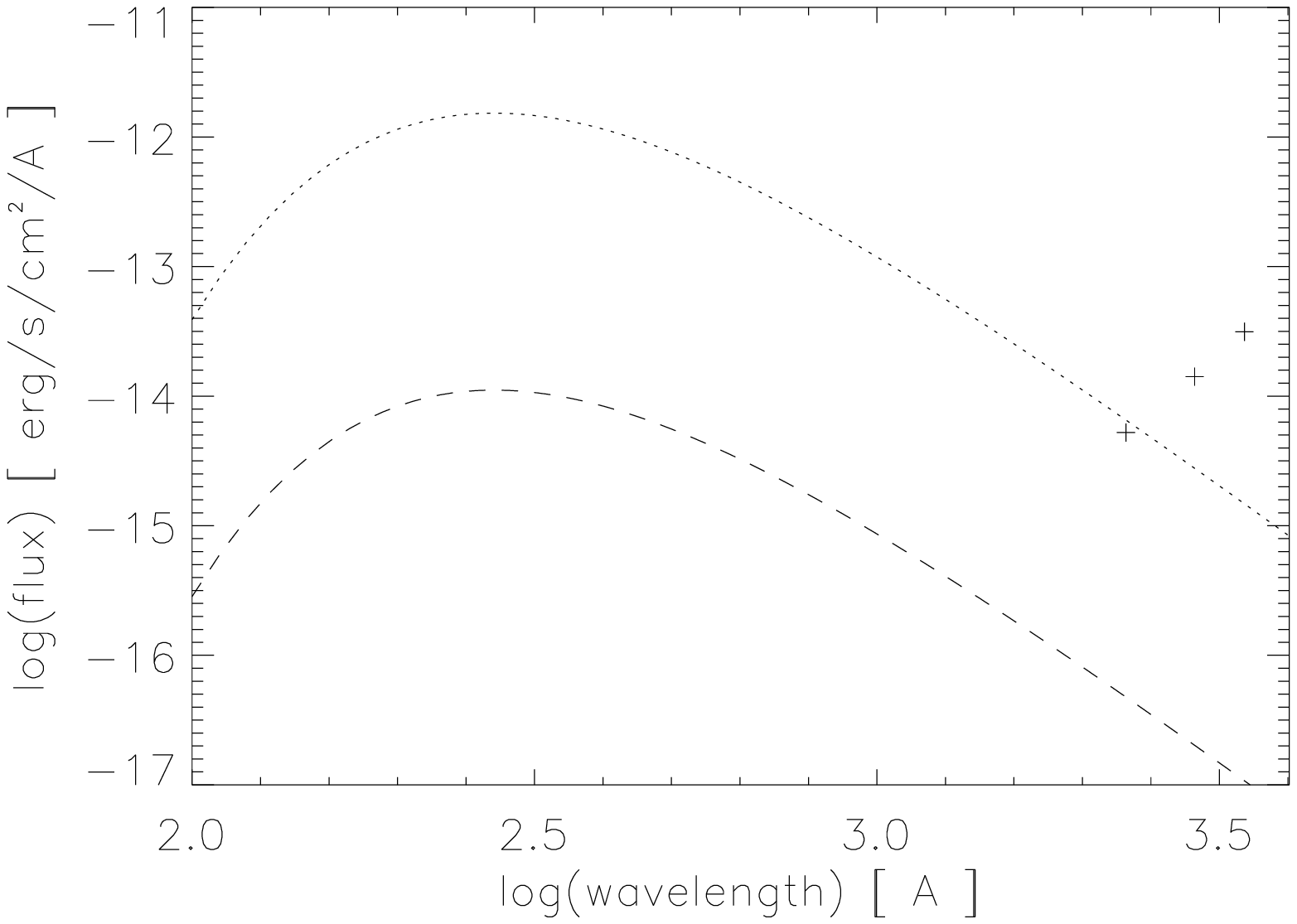}{Spectral energy distribution in the optical filters U, UVW1 and UVM2. Also plotted is blackbody emission representing a white dwarf with a temperature of 105000 K and radius of $8\times10^8$ cm at distances of 850 pc (dotted line) and 9.96 kpc (dashed line). The slope and flux level of the spectral energy distribution indicate that the optical flux is dominated by nebular emission.}

\subsection{Light curves}

We examined the OM photometry (Fig. 4). The optical light curves 
are consistent with a constant flux. The measured rms variations, those 
expected from Poisson statistics, and the ratios of these two quantities are 
listed in Table \ref{Tbl_rms}. $s$ and $s_{exp}$ are given in percentage of the 
mean value. The optical light curves in filters UVW1 and UVM2 where the ratio 
of measured to expected rms $s / s_{exp}$ is 1.27 and 1.21, respectively, may 
indicate weak variability, however, the presence of systematic and/or unkown 
errors is more likely. Since the optical flux is dominated by nebular emission, 
we do not expect variability in these bands.

\begin{table*}[!htb]
\caption{Measured and expected variations and their ratio}
\label{Tbl_rms}
\centering
\begin{tabular}{llll}
\hline\hline
optical filter & measured variation $s$ & 
expected variation $s_{exp}$ & ratio $s / s_{exp}$ \\
\hline
U    & 0.044 \% & 0.045 \% & 0.96 \\
UVW1 & 0.045 \% & 0.046 \% & 1.27 \\
UVM2 & 0.288 \% & 0.237 \% & 1.21 \\
\hline
\end{tabular}
\end{table*}

\IBVSfig{15cm}{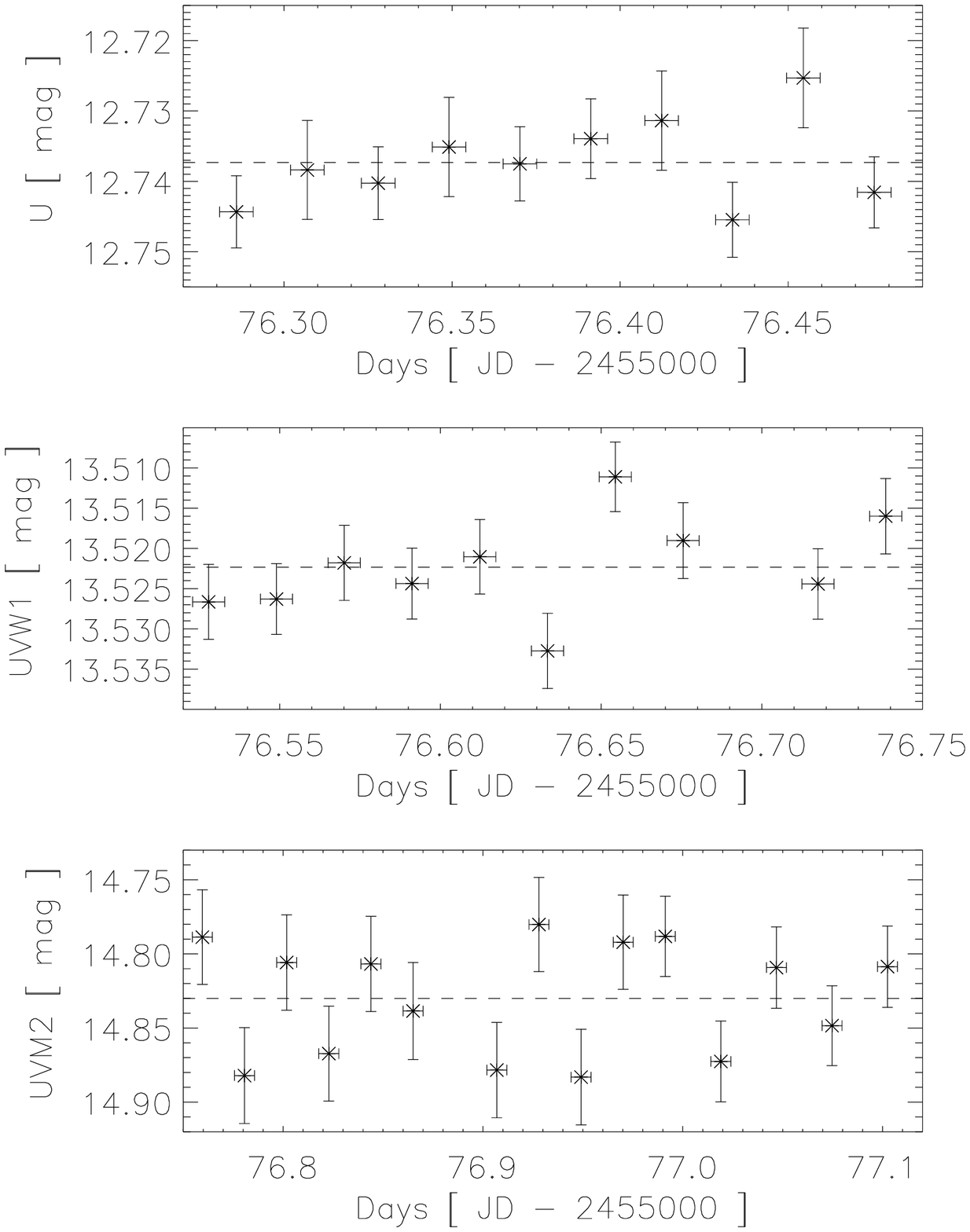}{Light curves of the OM exposures for filters U (top, 3440 \AA), UVW1 (middle, 2910 \AA) and UVM2 (bottom, 2310 \AA). The dotted lines show the average magnitude.}

\section{Discussion} \label{sec_dis}

We reported the non-detection of X-rays from the jet-driving symbiotic star 
HD\,149427 using {\em XMM-Newton}. However, we detected HD\,149427 with the 
Optical Monitor and extracted light curves in the filters U (3440 \AA), UVW1 
(2910 \AA) and UVM2 (2310 \AA). The light curves in filters UVW1 and UVM2 show 
hints for variability, however, due to possible systematic and/or unkown 
errors, they are also consistent with a constant flux. The UV spectral energy 
distribution indicate that the optical flux is dominated by nebular emission.

Using the effective exposure time of 33.5 ks in all instruments, the 
sensitivity of {\em XMM-Newton} (Watson et al. 2001) gives an upper limit of 
about 10$^{-15}$ erg s$^{-1}$ cm$^{-2}$ in the soft band ($< 2$ keV) and of about 
10$^{-14}$ erg s$^{-1}$ cm$^{-2}$ in the hard band ($> 2$ keV).

\subsection{What is the nature of HD\,149427?}

We can identify several different scenarios for the nature of HD\,149427 in the 
literature: a planetary nebula or a symbiotic star, both either close to us or 
very distant. 

Pereira et al. (2010) favor the idea of HD\,149427 being a distant PN at about 
10 kpc away from us. From the position of HD\,149427 in the 
[OIII]$\lambda$5007/H$\beta$ -- [OIII]$\lambda$4363/H$\gamma$ diagram, they 
claim it is neither consistent with that of PNe nor that of symbiotic stars. 
They argue that the electron density observed in HD\,149427 is higher than 
typical electron densities in PNe, but not as high as observed in D- and 
D'-type symbiotics. However, the closest object to HD\,149427 in this diagram 
is another D'-type symbiotic. Or is HD\,149427 a symbiotic star whose companion 
just turns into a PN?

In case of HD\,149427 being a central star of a PN, we simply expect X-ray 
emission from a blackbody with a temperature of about $10^5$ K 
(Gutierrez-Moreno \& Moreno 1998) as given in Fig. 3. The 
emission at 0.2 keV ($=$ 62 \AA ) would be fainter than our upper limits, so 
our non-detection would have no significant implications.

However, if HD\,149427 was a symbiotic star and accretion was present in this 
object, we would be able to test the accretion rate with important implications
(see below).

\subsection{What is the distance to HD\,149427?}

For deriving the distance to HD\,149427, Pereira et al. (2010) used their fits 
to high resolution spectra, namely their derived values of $\log\,g$ and 
$\log\,T$. They estimated the luminosity of the star and calculated the 
distance to it using a measured $V$ magnitude. They finally derived a distance 
of $9.94\pm0.6$ kpc. However, using the luminosity of the white dwarf in the 
system, the distance would result in a extremely large white dwarf radius of 
0.14 $R_\odot$ (Pereira et al. 2010). Turning this argument around and using 
radius determinations of 0.01--0.02 $R_\odot$ (e.g. Gutierrez-Moreno \& Moreno 
1998), we would get a distance of about 0.7--1.5 kpc. A second point of 
criticism is that the error estimate mentioned above bases only on the error of 
$\log\,g$ and $\log\,T$, but not on the error of the $V$ magnitude. The $V$ 
magnitudes in the literature vary between 11.2 and 12.7 -- the latter value has 
been used by Pereira et al. (2010). The former value, however, would imply a 
distance of only about 5 kpc, thus the distance based only on their own data is 
still highly uncertain. The smallest estimated distance to HD\,149427 is only 
850 pc (Gutierrez-Moreno \& Moreno 1998), therefore we adopt two extremal 
distances of 850 pc and 10 kpc.

With a distance of 850 pc, the flux upper limits give an upper limit for the
soft X-ray luminosity of about $8.6\times10^{28}$ erg s$^{-1}$ and for the hard 
X-ray luminosity of about $8.6\times10^{29}$ erg s$^{-1}$; with the highest 
claimed distance of 9.96 kpc, these limits increase to $1.2\times10^{31}$ erg 
s$^{-1}$ and $1.2\times10^{32}$ erg s$^{-1}$, respectively. 

\subsection{Implications for the accretion process}

Since accretion should be the main source of hard X-ray emission in this 
objects, we can convert these upper limits of luminosities into upper limits 
for the accretion rates. From the definition
\begin{eqnarray}
L_{\rm acc} &\ltsimeq& \frac{1}{2}\,\frac{G\,M\,\dot M}{R} = 3\times10^{32} 
\,\textrm{erg s$^{-1}$} \left(\frac{M}{0.6\,\textrm{M}_\odot}\right) \nonumber \\
&& \left(\frac{\dot M}{10^{-10}\,\textrm{M}_\odot\,\textrm{yr}^{-1}}\right)\,
\left(\frac{8\times10^8\,\textrm{cm}}{R}\right)\,
\end{eqnarray}
follows that, after assuming the largest distance, the white dwarf accretes 
with a rate below $10^{-10}$ M$_\odot$ yr$^{-1}$. At these low accretion rates, a 
boundary layer between the accretion disk and the surface of the white dwarf is 
optically thin (Pringle \& Savonije 1979; Popham \& Narayan 1995) and thus 
emitting hard X-rays with $L_{\rm acc}$. In case, the low distance of 850 pc is 
correct, the upper limit on the accretion rate reduces even to $10^{-13}$ 
M$_\odot$ yr$^{-1}$. 

\subsection{Implications for the jet}

Gutierrez-Moreno \& Moreno (1998) found an electron density of the potential 
jet knot of $10^5$ cm$^{-3}$, an electron temperature of 20000 K and a velocity 
of this knot of 120 km s$^{-1}$. With a typical ionization fraction at this 
temperature and a typical compression factor -- we assume that the observed 
values correspond to those of shocked jet material -- we derive an (un-shocked) 
jet density in the same order of magnitude as the measured electron density. 

The jet velocity is connected to the velocity of the contact discontinuity or 
knot via
\begin{equation}
v_{\rm knot} = v_{\rm jet}\,\frac{\sqrt{\eta}}{1 + \sqrt{\eta}}
\end{equation}
with $\eta$ being the ratio of jet to ambient density. Therefore the jet 
velocity is always higher than the knot propagation velocity (e.g. a factor 4.2
for $\eta = 0.1$ or 1.3 for $\eta = 10$). The presence of a jet with a velocity 
of about 120 km s$^{-1}$ naturally leads to shocks with temperatures of the 
order of 0.02 keV, which contribute only marginaly to the bands observed with 
{\em XMM-Newton} or {\em Chandra}. Higher velocities above 300 km s$^{-1}$, 
however, should lead to soft X-ray emission. Assuming a cylindrical jet with a 
radius of about 1 AU, the mass-loss rate in the jet is 
\begin{equation}
\dot M_{\rm jet} = \pi\,R_{\rm jet}^2\,n_{\rm jet}\,m_{\rm H}\,v_{\rm jet} > 
2.3\times10^{-11}\,\textrm{M}_\odot\,\textrm{yr}^{-1}\,.
\end{equation}
Since the mass-loss rate through the jet is at most a few percent of the mass
accretion rate (depending on the underlying jet formation model), the latter 
must have been of the order of $10^{-10}$ -- $10^{-8}$ M$_\odot$ yr$^{-1}$ at the 
time when the jet was formed. 

The kinetic luminosity of the jet would be 
\begin{equation}
L_{\rm kin} = \frac{1}{2}\,\dot M_{\rm jet}\,v_{\rm jet}^2 > 2\times10^{29}\,
\textrm{erg s$^{-1}$}\,.
\end{equation}
Only a fraction of this kinetic luminosity ($\sim 1$--20 \%, Stute \& Sahai 
2007) would be radiated away in soft X-rays, thus the presence of an X-ray 
emitting jet cannot be ruled out by our non-detection, independent of adopting 
the small or large distance. Furthermore we can argue that if a larger fraction 
of the proposed observation time had not been affected by high background, we 
might have detected soft emission from the jet.

Another source of soft X-ray emission might be the presence of colliding winds
as in $\beta$ systems classified by M\"urset et al. (1991) and M\"urset et al. 
(1997). Their typical luminosity may also be of the order of $10^{31}$ erg 
s$^{-1}$.

\section{Conclusion} \label{sec_concl}

We reported the non-detection of X-ray emission from the enigmatic object 
HD\,149427. This result poses upper limits on the mass accretion 
rate of the white dwarf. Such low accretion rates are untypical for symbiotic 
stars and may favor the picture of HD\,149427 being a young PN, even if we adopt
the larger values of its distance. Unfortunately, the distance to HD\,149427 is 
still highly uncertain.

We estimated the possible mass-loss rate and kinetic luminosity of the jet and
found no contradiction with our upper limit of soft X-ray emission. If a larger 
fraction of the proposed observation time had not been affected by high 
background, we might have detected soft emission from the jet. Therefore new 
X-ray observations might give new exciting results for this enigmatic object.

\vspace{1cm}

ACKNOWLEDGMENTS:
This work is based on observations obtained with {\em XMM-Newton}, an 
ESA science mission with instruments and contributions directly funded by ESA 
Member States and the USA (NASA). GJML thanks NASA for funding this work 
by XMM-Newton AO-8 award NNX09AP88G.

\references

Allen, D.A., Glass, I.S. 1974, MNRAS, 167, 337

Allen, D.A. 1982, in: IAU Colloquium 70 on ``The Nature of Symbiotic
stars'', p. 225, Friedjung M., Viotti R. (eds.), Reidel 

Brocksopp, C., Bode, M. F., Eyres, S. P. S. 2003, MNRAS, 344, 1264 

Brocksopp, C., Sokoloski, J. L., Kaiser, C., et al. 2004, MNRAS, 347, 430 

Galloway, D. K., Sokoloski, J. L. 2004, ApJ, 613, L61

Glass, I.S., Webster, B.L. 1973, MNRAS, 165, 77

Gutierrez-Moreno, A., Moreno, H. 1998, PASP, 110, 458

Gutierrez-Moreno, A., Moreno, H., Cortes, G. 1987, Rev. Mex. A. A., 14, 344

Gutierrez-Moreno, A., Moreno, H., Cortes, G. 1995, PASP, 107, 462

Henize, K.G. 1967, ApJS, 14, 125

Henize, K.G. 1976, ApJS, 30, 491

Karovska, M., Carilli, C. L., Raymond, J. C., Mattei, J. A. 2007, ApJ, 661, 
1048

Karovska, M., Gaetz, T. J., Carilli, C. L., Hack, W., Raymond, J. C., Lee, 
N. P. 2010, ApJ, 710, 132

Kellogg, E., Pedelty, J. A., Lyon, R. G. 2001, ApJ, 563, 151

Kellogg, E., Anderson, C., Korreck, K, et al. 2007, ApJ, 664, 1079

Kenny, H.T. 1995, PhD Thesis, University of Calgary

Kerber, F., Mignani, R. P., Smart, R. L., Wicenec, A. 2008, A \& A, 479, 155

Maciel, W. J. 1984, A \& AS, 55, 253

Milne, D.K. 1979, A \& AS, 36, 227

Milne, D.K., Aller, L.H. 1975, A \& A, 38, 183

Mukai, K., Ishida, M., Kilbourne, C., et al. 2006, PASJ, 59, 177

M\"urset, U., Nussbaumer, H., Schmid, H. M., Vogel, M. 1991, A \& A 248, 458

M\"urset, U., Wolff, B., Jordan, S. 1997, A \& A 319, 201

Nichols, J. S., DePasquale, J., Kellogg, E., et al. 2007, ApJ, 660, 651

Parthasarathy, M., Garcia-Lario, P., Pottasch, S. R., et al. 2000, A \& A, 355,
720

Peimbert, M., Costero, R. 1961, BOTT, 3, 33

Pereira, C. B., Baella, N. O., Daflon, S., Miranda, L. F. 2010, A \& A, 509, A13

Phillips, J.P. 2001, A \& A, 367, 967

Popham, R., Narayan, R. 1995, ApJ, 442, 337

Pringle, J. E., Savonije, G. J. 1979, MNRAS, 187, 777

Sokoloski, J. L., Kenyon, S. J., Espey, B. R., et al. 2006, ApJ, 636, 1002

Stute, M., Sahai, R. 2007, ApJ, 665, 698

Stute, M., Sahai, R. 2009, A \& A, 498, 209

Tajitsu, A., Tamura, S. 1998, AJ, 115, 1989

Watson, M. G., Augueres, J.-L., Ballet, J., et al. 2001, A \& A, 365, L51 

Webster, L. B. 1966, PASP, 78, 136

Wright, A.E., Allen, D.A. 1978, MNRAS, 184, 893
\endreferences

\end{document}